\documentclass[conference]{IEEEtran}
\usepackage{amsmath,amsfonts}
\usepackage{algorithmic}
\usepackage{algorithm}
\usepackage{array}
\usepackage[caption=false,font=normalsize,labelfont=sf,textfont=sf]{subfig}
\usepackage{textcomp}
\usepackage{stfloats}
\usepackage{url}
\usepackage{verbatim}
\usepackage{graphicx}
\usepackage{listings} 
\usepackage{multirow} 
\usepackage{enumitem}
\usepackage[section]{placeins}
\usepackage{needspace}
\usepackage{booktabs}
\usepackage[colorlinks,urlcolor=blue,linkcolor=blue,citecolor=blue]{hyperref}

\usepackage{color,array}

\usepackage{graphicx}

\begin{document}

\title{Large Language Model Integration with Reinforcement Learning to Augment Decision-Making in Autonomous Cyber Operations} 

\author{
\IEEEauthorblockN{
Konur Tholl\IEEEauthorrefmark{1},
Fran\c{c}ois Rivest\IEEEauthorrefmark{2},
Mariam El Mezouar\IEEEauthorrefmark{2},
Adrian Taylor\IEEEauthorrefmark{3},
Ranwa Al Mallah\IEEEauthorrefmark{4}
}

\IEEEauthorblockA{\IEEEauthorrefmark{1}
Department of Electrical and Computer Engineering\\
Royal Military College of Canada\\
Kingston, Canada}

\IEEEauthorblockA{\IEEEauthorrefmark{2}
Department of Mathematics and Computer Science\\
Royal Military College of Canada\\
Kingston, Canada}

\IEEEauthorblockA{\IEEEauthorrefmark{3}
Defence Research and Development Canada\\
Ottawa, Canada}

\IEEEauthorblockA{\IEEEauthorrefmark{4}
Department of Computer and Software Engineering\\
Polytechnique Montreal\\
Montreal, Canada}
}

\markboth{Journal of IEEE Transactions on Artificial Intelligence, Vol. 00, No. 0, Month 2025}
{Double-Blind Review}

\maketitle

\begin{abstract}
Reinforcement Learning (RL) has emerged as a promising approach for autonomous decision-making in the cybersecurity domain, enabling agents to learn through direct environment interaction. However, RL agents in Autonomous Cyber Operations (ACO) typically learn from scratch, requiring them to execute undesirable actions to learn their consequences. In this study, we integrate external knowledge in the form of a Large Language Model (LLM) pretrained on cybersecurity data that our RL agent can directly leverage during training to make informed decisions. By guiding initial training with an LLM, we improve baseline performance and reduce the need for exploratory actions with clearly negative outcomes. We evaluate our LLM-integrated approach in a realistic, simulated cybersecurity environment, and demonstrate that our guided agent achieves over 2x higher rewards during early training and converges to a favorable policy approximately 4,500 episodes faster than the baseline.
\end{abstract}


\section{Introduction}
\label{intro}
Offensive cyber operations have increased at an unprecedented scale over recent years \cite{forbesCyberStats, tholl_thesis_2025, martin_cyberstats24-25_2025, canada_nationalcyberransomware_2024}. It is no longer feasible for humans to manually defend their IT systems, motivating the development of automated tools to help with the intrusion detection process \cite{chakraborty_intrusion_2013}. These tools began as signature-based approaches, which require known attacks to exist in a database to be successfully detected \cite{saeed_review_2024}. Because an exact match is required for detection, these systems are vulnerable to any new exploits whose signatures are not yet stored.  

Machine Learning (ML) approaches helped overcome these deficiencies by training models to detect patterns across malicious and benign samples, eliminating the requirement for exact matches \cite{center_for_security_and_emerging_technology_machine_2021}. However, traditional supervised approaches depend on large, static datasets to train models. These datasets must be manually updated whenever new behaviors emerge; a task that is not scalable given the rapidly evolving nature of cybersecurity.

Autonomous Cyber Operations (ACO) is an emerging field in which specialized cybersecurity agents are trained to make effective decisions on behalf of humans \cite{baillie_cyborg_2020}. Unlike traditional ML, ACO employs Reinforcement Learning (RL), where agents can be trained through direct interaction with their environment, eliminating the reliance on large, static, labeled datasets \cite{sutton_reinforcement_2014}. This interaction enables agents to dynamically adapt to evolving adversarial Tactics Techniques and Procedures (TTPs), offering a scalable and adaptive approach to modern cyber defense.

For ACO to operate effectively, an environment must exist that mimics cybersecurity, and produces the necessary signals for an agent to learn. The Technical Cooperation Program (TTCP) has created such an environment: CybORG, a widely used RL framework designed to train agents to defend a simulated enterprise network against adversarial threats \cite{baillie_cyborg_2020, cagechallenge2github}.

Despite this progress, RL agents in ACO typically begin as ``bare-bone'' models, requiring substantial training time before converging to an optimal policy \cite{wiebe_learning_2023, mcdonald_competitive_2024}. Moreover, because the agents' weights are initially randomized, they are equally likely to select favorable or unfavorable actions. This poses significant risks in the context of cybersecurity, where an incorrect action can lead to the compromise of entire IT systems. 

In addition, the data involved in the incident detection and response process is typically textual by nature (e.g., log files) and must be mapped to a numerical feature space for RL agents to process. This feature engineering process is typically deterministic, and could miss critical information relating to  cybersecurity incidents \cite{baillie_cyborg_2020}.

These limitations can be mitigated by incorporating external knowledge into the RL pipeline that agents can directly leverage during decision-making. Instead of acting randomly and learning solely from environmental feedback, an agent can benefit from a teacher that provides guidance, allowing continued learning from a strong baseline \cite{pfeiffer_reinforcedimitationlearning_2018}. Such a teacher should be generalizable across cybersecurity environments, and capable of effectively identifying patterns in the textual information present in cybersecurity data.

In this study, we integrate a Large Language Model (LLM) into the ACO pipeline to improve an RL agent's decision-making capabilities. The vast amounts of textual data used to train LLMs make them ideal candidates to serve as a teacher, enabling them to generalize across cybersecurity environments and capture contextually relevant information that might otherwise be missed during feature engineering \cite{zhou_largellm4teach_2024}. 

\hyphenation{teacherguided}
In particular, the major contributions this study makes are:
\begin{itemize}
    \item \textit{LLM Integration for Efficient RL Training}. We demonstrate that incorporating an LLM into the RL pipeline improves training efficiency and reduces the need for suboptimal exploratory actions. Specifically, our guided agent converges to a favorable policy approximately 4,500 episodes earlier than the baseline and receives over twice the rewards during the early stages of training.
    \item \textit{A Novel Teacher-Guided Technique}. We introduce a new method that incorporates LLM feedback using both action masking and an auxiliary loss signal.
    \item \textit{A Framework to Evaluate LLMs}. We propose a methodology to efficiently assess the incident response capabilities of LLMs pretrained for cyber defense. 
\end{itemize}

\textit{Organization}. The rest of this paper is organized as follows:
\begin{itemize}
    \item The \textit{Related Work} section provides an overview of LLMs, RL, and cybersecurity, and discusses previous work in these fields.
    \item The \textit{Methodology} section describes the design and implementation of our research activities and experimental setup. 
    \item The \textit{Evaluation} section presents our results and explains the rationale behind major implementation decisions.
    \item  Finally, the \textit{Conclusion} summarizes our contributions, discusses limitations, and outlines directions for future work.
\end{itemize}

%Commented out for double-blind review
% Portions of this work are adapted from the author's Master's thesis, submitted to the Royal Military College of Canada (RMC) \cite{tholl_thesis_2025}.
 
\section{Related Work}
\label{background}
In this section, we review existing studies in related fields and synthesize insights to propose a novel approach for enhancing current applications in ACO. Background information on the related fields is described at a high level.

\subsection{LLMs and RL}
RL is a distinct Machine Learning (ML) paradigm, where an agent learns by interacting with its environment through direct action execution \cite{sutton_reinforcement_2014}. The actions produce a new state and a reward, which the agent uses in a trial-and-error like fashion to learn a strategy (i.e., a policy) for selecting the ones that maximize the cumulative reward.

One consideration for using traditional RL is that agents start as bare-bones models that must be trained from scratch. This is inefficient because initially, the agent is just as likely to choose an obviously unfavorable action as a favorable one. As a result, it will inevitably make poor decisions before it can learn to select optimal ones, increasing the training time required to converge to a favorable policy. 

The idea of having the agent leverage a teacher to improve training efficiency is not novel in itself, and many previous studies have used different teacher-guided techniques; however, these typically require a pretrained agent to act as a teacher, which necessitates additional training \cite{beikmohammadi_ta-explorerewardshaping_2023, pfeiffer_reinforcedimitationlearning_2018, wang_learningactionmasking_2024}. Furthermore, the features that are provided to these agents must be manually engineered, which can be problematic in data-rich environments like cybersecurity, where vital information may be missed in the feature engineering process.

A potential solution is to have an LLM act as the teacher. Fundamentally, an LLM is a Deep Neural Network (DNN) designed to recognize patterns in language and produce contextually relevant responses \cite{naveed_comprehensive_2024}. We primarily use decoder-only LLMs in this study, which auto-regressively generate responses token by token \cite{roberts_how_2024, openai_gpt-4_2024}.

Textual environment data can be fed directly into the LLM, alleviating concerns about missing important information due to manual feature engineering. Moreover, an LLM's vast generalized knowledge eliminates the need to train an additional agent to act as the teacher. 

Because the LLM's output is textual, it must be mapped into an executable action for the environment. W. Huang et al. proposed a solution to this problem in which they used an LLM to perform actions in the VirtualHome environment, which simulates basic household activities \cite{huang_language_2022}. They leveraged RoBERTa, an encoder-only type model, to convert the LLM's textual output into executable actions; however, they did not explore leveraging the LLM in the RL process \cite{yinhan_roberta_2019}, where an agent can learn and adapt to particular environments. 

This lack of learning is addressed in the solution proposed by M. Kwon et al., which uses GPT-3, a decoder-only LLM, to compute rewards for the RL process \cite{kwon_llmrewardshaping_2023, tom_language_2020}. This method works by appending the environment's objectives to the prompt, and having the LLM evaluate whether they align with the outcome of the episode, increasing the reward signal if they do. This approach requires the agent to wait an entire episode before receiving any feedback from the LLM. Moreover, this method assumes the environment's objectives are fixed for an entire episode; however, in more complex environments, the goals may change based on the state, making such assumptions unrealistic.

Z. Zhou et al. proposed \textit{LLM4Teach}, a method in which a pretrained LLM guides an RL agent's training \cite{zhou_largellm4teach_2024}. The RL algorithm employed is Proximal Policy Optimization (PPO), an actor-critic method that outputs a probability distribution over possible actions. The LLM's guidance is incorporated as an additional loss signal for the actor - in particular, they propose \cite{zhou_largellm4teach_2024}:

\begin{equation}
    \label{llm4teachequation}
    \mathcal{L}(\theta) = \mathcal{L}_{\text{RL}}(\theta) + \lambda \mathbb{E}_{s \sim \pi_{\theta}} \mathcal{H} \left( \pi_T(\cdot|s) \| \pi_\theta(\cdot|s) \right)
\end{equation}

where \(\mathcal{L}_{RL}(\theta)\) is the standard PPO loss, and the term \(\mathbb{E}_{s \sim \pi_{\theta}} \mathcal{H} \left( \pi_T(\cdot|s) \| \pi_\theta(\cdot|s) \right)\) 
represents the deviation between the RL agent's and the teacher's policy. The weighting coefficient \(\lambda\) is gradually decayed, reducing the LLM's influence over time.

To generate a distribution over actions, the LLM was queried once for each possible action, which is inefficient for environments with large action spaces. The authors acknowledged this limitation and suggested directly extracting the model's logits, which would require only a single forward pass. 

While each of the discussed studies provides a valuable contribution, they have implemented the teacher-guided techniques (e.g., reward shaping) in isolation of each other. This presents a great opportunity to combine complementary approaches for incorporating a teacher's guidance.

Furthermore, the LLMs leveraged in prior RL research were trained on a corpus of generic data, rather than the specific domain in which they are deployed. Incorporating an LLM that has been pretrained on a particular domain could yield superior performance in terms of reward signals and training efficiency.

\subsection{LLMs and ACO}
Previous work has demonstrated the potential benefits of integrating LLMs into the cyber domain, showing success in tasks ranging from Distributed Denial of Service (DDoS) detection to providing transparent explanations in the incident response process \cite{guastalla_applicationllmddos_2024, loevenich_designllmcyborg_2024, ali_huntgpt_2023}. However, there is no active learning in this work - the recommended actions are based solely on the static, pre-existing datasets used to train the LLM. This lack of continuous learning makes the systems susceptible to adversary behavior whose patterns are not captured by the LLM's training data. Incorporating the LLM into the ACO process using teacher-guided techniques similiar to the ones discussed above enables systems to dynamically adapt to new threats in a scalable manner, while leveraging the LLM's vast knowledge to make effective early decisions.

ACO refers to conducting cybersecurity tasks with minimal or no human intervention \cite{farooq_generic_2024}. The goal of ACO is not to replace cyber operators, but to work alongside them to improve the overall effectiveness of cybersecurity. Modern ACO applications use RL to train agents, necessitating an environment in which agents can interact.

Previous environments have been created to replicate the cyber domain, but many were either not scalable or not designed for RL, which requires sufficiently informative signals to guide training \cite{baillie_cyborg_2020}. These limitations were addressed in CybORG's Cage Challenge 2, a simulated environment that mimics cybersecurity and provides frequent feedback, enabling stable policy learning \cite{kiely_autonomous_2023}.

\subsection{Discussion}
There has been substantial progress in integrating RL into ACO; however, current approaches require agents to learn from scratch, leading to longer training times and suboptimal initial policies \cite{bhagyalakshmi_machine_2024, baillie_cyborg_2020, kiely_autonomous_2023, mcdonald_competitive_2024, wiebe_learning_2023, farooq_generic_2024}. Prior studies have demonstrated that guiding the RL process using an LLM can alleviate these concerns, but none have been directly applied to the cyber domain \cite{pfeiffer_reinforcedimitationlearning_2018, zhou_largellm4teach_2024, wang_boostinginstruccomprehension_2025}.   

Despite the promising overlap of LLMs, RL, and ACO, limited research has been conducted in this area. J. Loevenich et al. integrated an LLM into the CybORG environment, but this was used to increase transparency for the end user regarding the RL agent's decisions. The LLM had no impact on the training process \cite{loevenich_designllmcyborg_2024, baillie_cyborg_2020}.

This presents a valuable opportunity to explore integrating an LLM into the RL pipeline for ACO, with the primary objective of augmenting decision-making. Additionally, the discussed teacher-guided techniques have largely been studied in isolation from each other, creating an opportunity to combine complementary approaches to optimize training (e.g., using an auxiliary loss signal alongside action masking).

\section{Methodology}
\label{methodology}
In this section, we outline the phased approach we used to integrate an LLM into the RL process to augment decision-making in ACO. These phases include:

\begin{enumerate}[label=\textit{\Alph*.}]
    \item \textit{LLM Evaluation}. Existing decoder-only LLMs, pretrained on cybersecurity were systematically reviewed using a dataset of predefined CybORG-specific question-answer pairs, enabling us to identify the most suitable model.
    \item \textit{Teacher-Guided Techniques}. Various combinations of teacher-guided algorithms were implemented to efficiently incorporate the LLM into the training process.
    \item \textit{LLM Integration}. The LLM that performed the best in Phase 1 was incorporated into the RL pipeline. This involved:
    \begin{enumerate}[label=\arabic*)]
        \item Refining the prompt beyond what was used in Phase 1 during LLM selection.
        \item Implementing a method to reliably extract an executable action from the LLM's response.
        \item Integrating the LLM into the decision-making process using the best-performing teacher-guided technique.
    \end{enumerate}
\end{enumerate}

\subsection{LLM Evaluation}
\label{meth:llmselection}
Our evaluation comprised of four open-source, decoder-only LLMs that were pretrained on cybersecurity-related data. We chose these models based on availability, baseline architecture, and existing reviews. This evaluation process involved validating the performance of these LLMs on a dataset of cybersecurity-related question-answer pairs. We present the LLMs involved in the evaluation in Table \ref{tab:llmsEvaluatedPhase1}.

%!b to make sure this appears after its mentioned in the text
\begin{table}
    \centering 
    \caption[LLMs Involved in the Evaluation.]{LLMs involved during the evaluation process.}
    \resizebox{1\columnwidth}{!} {
      \begin{tabular}{lcc}
        \toprule
        \textbf{LLM Name} & \textbf{Baseline Architecture} & \textbf{Source} \\
        \midrule
        Cyberdost2B \cite{huggingface_cyberdost} & Navarasa-2.0-2B & Hugging Face \\
        Lily7B \cite{huggingface_lily} & Mistral-7B & Hugging Face \\
        Cyber-Risk-Llama8B \cite{huggingface_cyber-risk-llama} & Meta-Llama-8B & Hugging Face \\
        Z7sec \cite{huggingface_z7sec} & Zephyr-7B & Hugging Face \\
        \bottomrule
      \end{tabular}
    }
    \label{tab:llmsEvaluatedPhase1}
\end{table}

\begin{figure}
\centering
    \centering
    \includegraphics[width=\columnwidth]{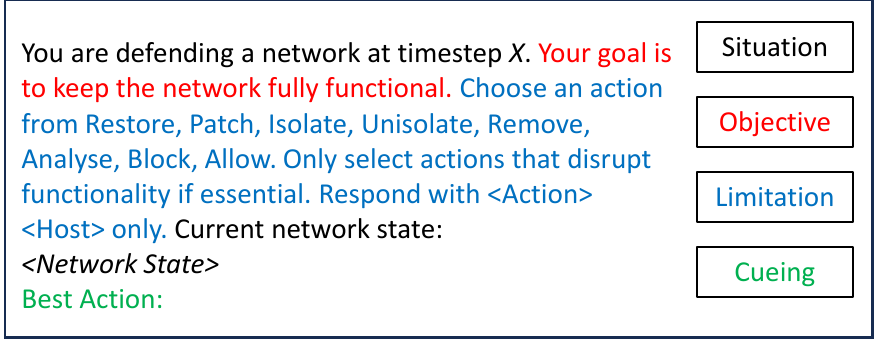}
    \caption[Prompt Design.]{Initial prompt design for evaluating LLMs. For clarity, the components of the prompt are color-coded.}
    \label{fig:initialPromptDesign}
\end{figure}

To ensure the LLMs were provided with input that could be effectively analyzed to generate a relevant response, we performed standard prompt engineering based on existing best practices \cite{ekin_prompt_2023, matthew_prompt_2024}. This evaluation focused on the LLMs' baseline knowledge, requiring prompts to be generic rather than task-specific. For example, the definition of an action in the context of CybORG was not included in the prompt - the evaluation aimed to assess the LLMs' pretrained capabilities. To ensure an unbiased assessment, we used an identical prompt format across all evaluated LLMs, as illustrated in Fig. \ref{fig:initialPromptDesign}.

Following the prompt structure shown in Fig. \ref{fig:initialPromptDesign}, we created questions of varying difficulty to evaluate the chosen LLMs. The difficulty was determined by the number of hosts involved:
\begin{itemize}
    \item Easy: 1-2 hosts.
    \item Medium: 3-7 hosts.
    \item Hard: 8-13 hosts.
\end{itemize}

We developed a Python script to generate a diverse set of these questions, enforcing predefined rules to ensure that the constraints of the CybORG environment were respected. For example, the operational server would not be compromised at timestep 0. These scenarios were converted into condensed JavaScript Object Notation (JSON) and sentence formats to support pattern recognition by the LLMs. In total, we generated 100 questions: 20 easy, 40 medium, and 40 hard.

To obtain the ground truth answers, we manually selected the most contextually relevant action from CybORG's predefined action space for each question. Emphasis was placed on selecting the action that would yield the highest reward for the episode, rather than optimizing for a single timestep.

To support the evaluation of the LLMs, we developed a script that compared their responses to the manually selected answers. This comparison used BERTScore, an encoder-only model that computes semantic similarity between text pairs based on contextual relevance \cite{zhang_bertscore_2020}. Similarity is measured by token-level precision, recall, and F1, which balances precision and recall. 

%Moving down in text so don't have this figure immediately proceed a table

We prioritized precision, as it more strongly penalizes irrelevant tokens. For example, if the correct answer is ``Isolate HostA'', and the LLM outputs ''Isolate, Patch, Restore, Remove, HostA HostB HostC``, recall would remain high because relevant tokens are included, but precision would drop due to excessive unrelated content. Encouraging this behavior is problematic, as it prevents the extraction of a single, concrete action from the LLM's output, making it inefficient for decision-making.

\begin{figure*}
    \centering
    \includegraphics[width=0.8\textwidth]{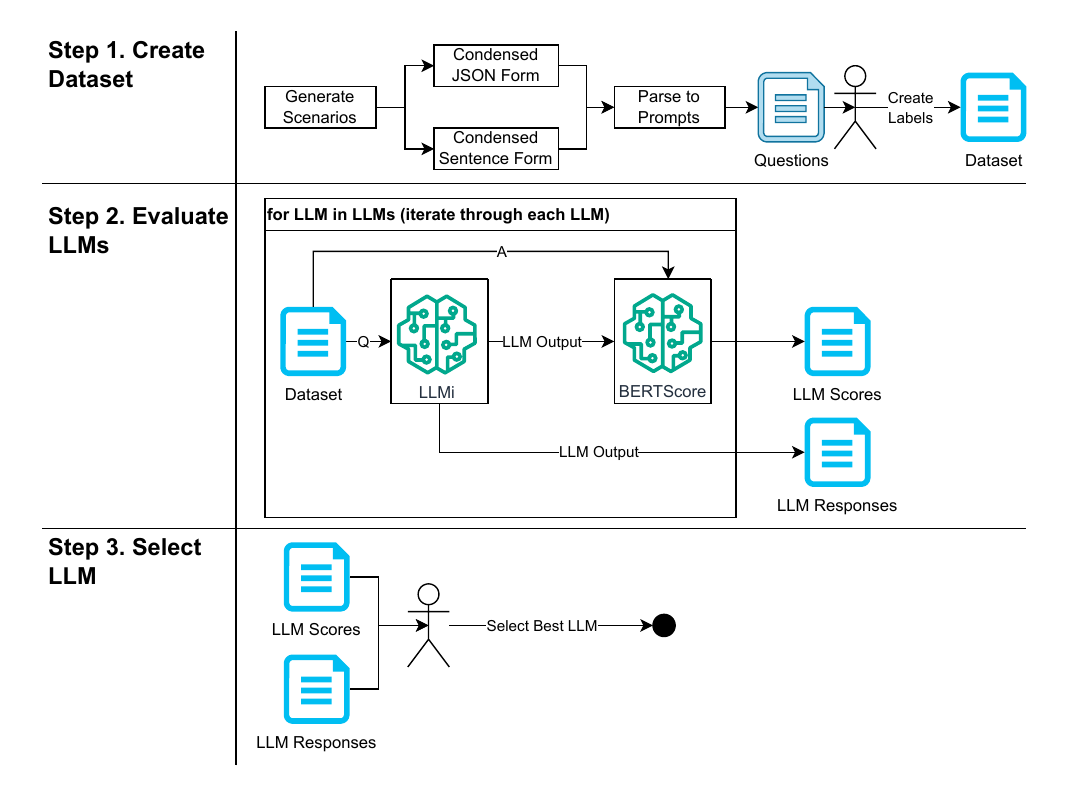}
    \caption[LLM Selection Overview.]{Overview of the process used to select an LLM. 
    Step 1 involved creating a dataset of questions and corresponding answers to evaluate the LLMs.
    In Step 2, the LLMs' predictions were recorded and evaluated against this dataset using BERTScore to compute precision, recall, and F1 scores.
    In Step 3, the results from Step 2 were manually reviewed, and the best-performing LLM was selected.}
    \label{fig:selectingBestLLM}
\end{figure*}

We recorded the LLM's textual output for both the JSON-based and sentence-based questions, alongside BERTScore's metrics to support manual validation. We present the entire procedure for selecting an LLM in Fig. \ref{fig:selectingBestLLM}.

%Insert figure for selecting best LLM here

\subsection{Teacher-Guided Techniques}
\label{meth:teacherguidedtechniques}
We implemented various combinations of existing teacher-guided techniques to identify the best approach for integration with the CybORG environment \cite{kwon_llmrewardshaping_2023, wang_learningactionmasking_2024, wang_boostinginstruccomprehension_2025, zhou_largellm4teach_2024, thollteacherguidedevaluation2025}. These combinations were informed by the individual techniques evaluated in the study by Tholl et \textit{al.} \cite{thollteacherguidedevaluation2025}. Unless otherwise noted, we implemented each component as described in \cite{thollteacherguidedevaluation2025}. 

Due to an LLM's substantial computational requirements, we used a pretrained RL agent - which is significantly less resource-intensive - to act as the teacher, enabling a more efficient evaluation. Moreover, because the teacher agent was trained on the same environment, issues attributable to the LLM (e.g., outputs that fail to conform to CybORG's environmental constraints) could be ruled out, allowing us to identify the most effective integration method. 

We then selected the best-performing combination in terms of training efficiency and final policy performance as the technique for LLM integration. In particular, we implemented the following combinations:
\begin{enumerate}
    \item Feature space modification with reward shaping.
    \item Action masking with feature space modification.
    \item Action masking with auxiliary loss.
\end{enumerate}

To maintain a focused scope, we used PPO as the baseline RL agent, applying the same parameters as in \cite{thollmodifycyborg2025}.

\subsubsection{Feature Space Modification with Reward Shaping}
As shown in Tholl et \textit{al.}'s study, incorporating reward shaping and feature space modification independently yields no noticeable improvements in training efficiency \cite{thollteacherguidedevaluation2025}. In contrast, our motivation for combining these two approaches was to provide the agent with an incentive to select the teacher's recommendation, thereby increasing its ability to map that recommendation to an executable action.

We employed the same one-hot encoding method as \cite{thollteacherguidedevaluation2025}, in which the teacher's recommended action was mapped to a vector, where the element at the action index was set to 1, while all other elements were set to 0. 

We then appended the one-hot encoded recommendation to the RL agent's state:

\begin{equation}
    s_{t}=[s_{t}^{\text{init}},\text{onehot}(a_{t}^{\text{Teacher}})]
\end{equation}

where \( s_{t}^{\text{init}} \) is the initial state returned from CybORG, and \(\text{onehot}(a_{t}^{\text{Teacher}})\) is the one-hot encoded teacher recommendation.

We also modified the agent's reward signal using the same recommendation:
\begin{equation}
\label{eq:featureandrewardshaping}
r_{t}(a)=
\begin{cases} 
      r_{envt}(a) + c_{1}, & \text{if } a = a_{t}^{\text{Teacher}}  \\
      r_{envt}(a), & \text{otherwise}
\end{cases}
\end{equation}

where \(c_{1}\) is a scalar that decays at each training interval (8 episodes), thereby reducing the teacher's impact on the reward signal.

\subsubsection{Action Masking with Feature Space Modification}
This combination is similar to the previous technique, but it forces the agent to select the teacher's recommendation. The direct influence of the masking is intended to help the agent map the teacher's recommendation to the corresponding action. We used the same one-hot encoding technique to map and append the teacher's recommendation to the agent's feature space. As in the preceding combination, the individual techniques used were taken from \cite{thollteacherguidedevaluation2025}.

Specifically, we employed inference-only action masking, where the masking affected only the agent's action selection and was not directly applied during training. In particular, the masking mechanism was implemented as follows:

\begin{equation}
\pi_{masked \theta}(a_{t})=\pi_{\theta}(a_{t}) * M_{t}(a_{t})
\end{equation}

where \(\pi_{masked \theta}(a_{t})\) is the masked policy and \(\pi_{\theta}(a_{t})\) is the original policy. The masking \(M_{t}(a)\) is defined as:

\begin{equation}
\label{eq:featureandactionmasking}
M_{t}(a)=
\begin{cases} 
      1, & \text{if } a=a_{t}^{\text{Teacher}} \\
      c_{2}, & \text{otherwise}
\end{cases}
\end{equation}

where the value of \(c_{2}\) was gradually increased, to facilitate a smooth transition from teacher-guided learning to independent RL. To ensure the masked policy remained a valid probability distribution, we normalized it so that all probabilities summed to 1:

\begin{equation}
    \pi_{masked \theta}(a_{t}) = \frac{\pi_{masked \theta}(a_{t})}{\sum_{a} \pi_{masked \theta}(a)}
\end{equation}

We used the masked distribution, \(\pi_{masked \theta}(a_{t})\), solely for action selection; it was not directly used to compute the actor's loss.

\subsubsection{Action Masking with Auxiliary Loss}
Tholl et al.’s study showed that masking actions during inference yielded initial gains in performance, with a subsequent dip during the transition from teacher-guided to independent RL \cite{thollteacherguidedevaluation2025}. In contrast, incorporating feedback via an auxiliary loss exhibited poor initial performance, but then quickly converged to the teacher’s policy. Our goal in combining action masking and auxiliary loss was to retain the early performance gains offered by action masking, while using the auxiliary loss signal to facilitate a smoother transition to independent RL. Similar to the above, we applied inference-only masking; masking was used exclusively for sampling actions, while unmasked probabilities were used to compute the actor's loss. 

We defined the auxiliary loss signal as the log probability of selecting the teacher's recommended action under the agent's policy:

\begin{equation}
    L_{teacher} = -log\pi_{\theta}(a_{t}^{Teacher}|s^{t})
\end{equation}

We then added the teacher loss, \(L_{teacher}\) to the actor's loss, scaling each by \(1-\sigma\) and \(\sigma\), respectively \cite{thollteacherguidedevaluation2025,zhou_largellm4teach_2024,beikmohammadi_ta-explorerewardshaping_2023}:

\begin{equation}
\label{eq:actionmaskingandauxloss}
    L_{tot}(\theta)=\sigma*L_{A}(\theta) + (1-\sigma) * L_{Teacher} + c_{3}S(\pi_{\theta}(\cdot|s_{t}))
\end{equation}

where \(\sigma\) was gradually increased at each training interval, allowing for a smooth transition from teacher-guided to independent RL. \(L_{A}(\theta)\) is the standard PPO loss for the actor network \cite{schulman_proximal_2017}. \( S(\pi_{\theta}(\cdot|s_{t})) \) represents the agent's entropy - the randomness within its policy - and \(c_{3}\) is the entropy coefficient, where a higher value encourages exploration.

During the teacher-guided phase, we incremented the entropy coefficient, \(c_{3}\) at each training interval to encourage the agent to explore actions beyond the teacher's recommendations, helping it surpass the teacher. After transitioning to independent RL, we reduced the entropy coefficient, favoring exploitation as the agent's policy began to stabilize.

\subsection{LLM Integration}
\label{meth:llmintegration}
In this subsection, we describe how the LLM was incorporated into the RL pipeline to enhance decision-making.

\subsubsection{Prompt Design}
To improve the quality and consistency of responses, we refined the prompt beyond what was used for the LLM evaluation, tailoring it for the selected LLM.

To begin, we replaced the CybORG-specific names with generic labels for both actions and hosts (e.g., action1, host1). Our decision was motivated by an observed inconsistency between the LLM's behavior and CybORG's terminology. For example, despite explicit instructions, the LLM often prioritized hosts labeled ``enterprise'' over the operational server, suggesting it relied more on training biases rather than the prompt's logic.

Secondly, we represented hosts' priorities as their distance from the operational server. Specifically, we parsed the scenario's Yet Another Markup Language (YAML) configuration file and applied a breadth-first search from the operational server to compute the minimum number of hops for each host along their respective critical paths. An example of how this priority encoding was included in the prompt is shown below (where 0 represents the highest priority):
\begin{verbatim}
{`Op_Server0': 0, 
`Enterprise2': 1, 
`Enterprise1': 2, `Enterprise0': 2, 
`User1': 3, `User2': 3, `User3': 3, 
`User4': 3}
\end{verbatim}

\subsubsection{Extracting LLM Recommendations}
\begin{figure}
    \centering
    \includegraphics[width=\columnwidth]{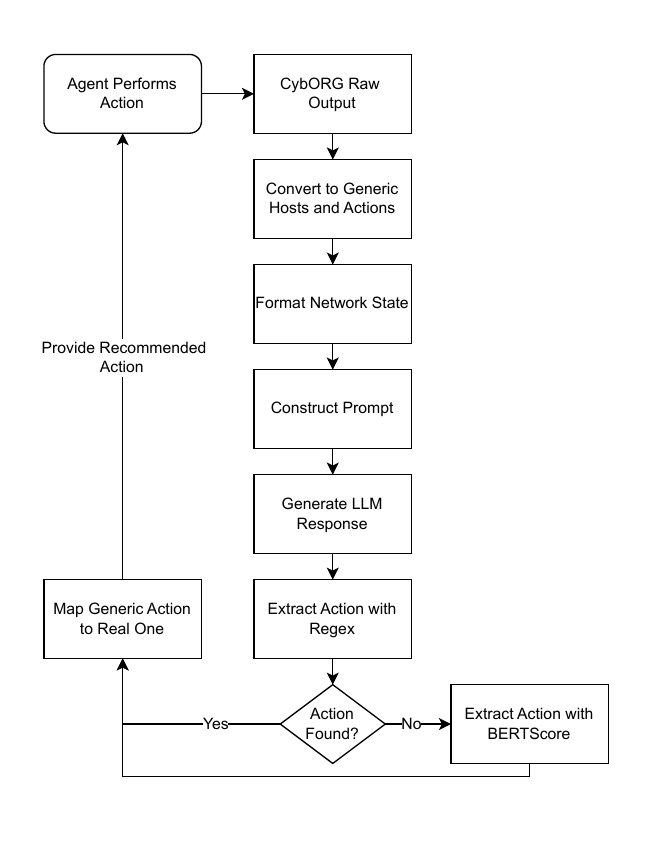}
    \caption[Flowchart for LLM Interaction.]{Overview of transforming CybORG's raw state into a coherent prompt, generating a response with the LLM, and extracting the corresponding action.}
    \label{fig:extractingLLMResponseFlowDiagram}
\end{figure}

As decoder-only LLMs output tokens instead of probability distributions over actions, their textual outputs must be mapped to executable actions \cite{roberts_how_2024}. To achieve this, we applied regular expressions (regex) to extract the first valid host and action from the LLM's response. If either a host or an action was not identified via regex, we used BERTScore to compute precision scores between the LLM's response and all possible hosts and actions. We then selected the host and action with the highest precision scores as the recommended executable action. Fig. \ref{fig:extractingLLMResponseFlowDiagram} illustrates the process of mapping CybORG's raw state into an engineered prompt for the LLM and mapping the LLM's response into an executable action.

\subsubsection{Integrating the LLM}
\begin{figure*}
    \centering
    \includegraphics[width=1.5\columnwidth]{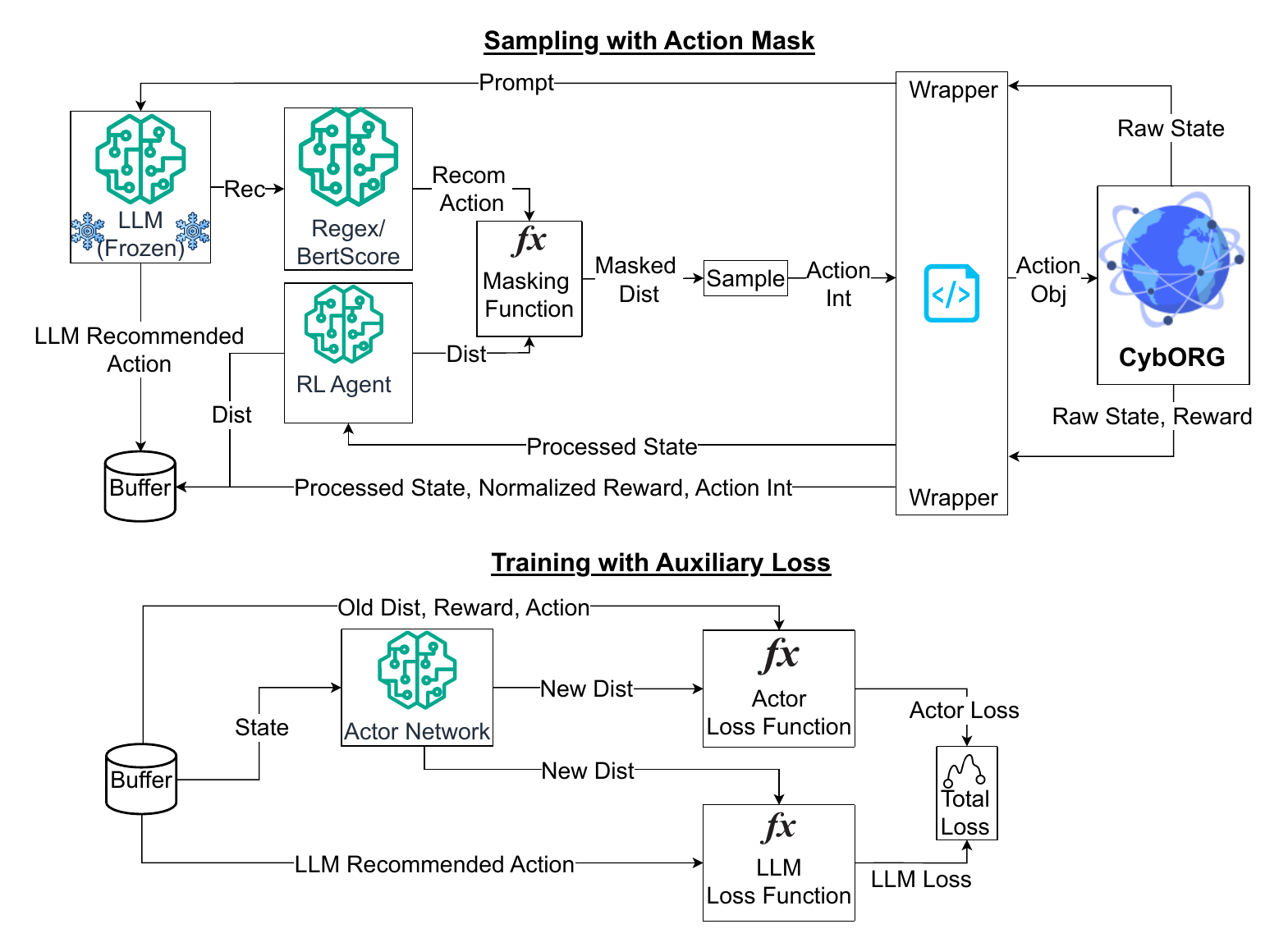}
    \caption[LLM Integration Design.]{Diagram illustrating the integration of the LLM into the RL pipeline. The LLM's guidance is applied using action masking at inference and as an auxiliary loss signal during training. \textit{Frozen} indicates that the LLM's parameters remain unchanged throughout training. To keep the diagram concise, the critic network is omitted, and some terms are presented in an abbreviated form.}
    \label{fig:integratingLLMwithAuxLossAndActionMask}
\end{figure*}

After establishing a robust method to reliably map the LLM's output into an executable action, we integrated it into the RL pipeline using the action masking and auxiliary loss combination. A diagram showing this integration is provided in Fig. \ref{fig:integratingLLMwithAuxLossAndActionMask}.

We provide the final hyperparameters for this integration in Table \ref{tab:finalHyperparametersLLMIntegration}.

\subsection{Evaluation Design}
We evaluated the effectiveness of the LLM-guided agent by measuring its performance using CybORG's reward signals and tracking the number of timesteps required to converge to a favorable policy. All evaluations were conducted using CybORG's default Cage Challenge 2 scenario, which simulates a 13-host, 3-subnet network \cite{cagechallenge2github}. 

We used Standard Error (SE) as the metric to quantify variability around the mean \cite{altman_standarddevanderror_2005}. We chose SE over Standard Deviation (SD) because our focus was to compare variance across algorithms, rather than dispersion between individual runs, but both are valid measurements of variability. 

We also used Local Interpretable Model-agnostic Explanations (LIME) to assess the LLM's impact on training.

\subsubsection{Quantifying the Teacher's Influence}
Following prior work, we used LIME to evaluate the impact of the teacher-provided features on the agent's decisions \cite{Faizan_MARL_2024, thollteacherguidedevaluation2025, ribeiro_lime_2016}. LIME functions by training an explainable model to estimate the contribution of each feature to an agent's selected action, where higher values indicate greater influence.

As in \cite{thollteacherguidedevaluation2025}, we initialized the LIME explainer with 500 observations sampled from CybORG to define the feature distributions. We then used perturbations of a selected state to estimate each feature's influence on the agent's action. To ensure an unbiased evaluation, we used the same base state for both feature space modification combinations \cite{thollteacherguidedevaluation2025}.

\begin{table}
    \centering
    \caption[Hyperparameters for LLM Integration.]{Final hyperparameters used for the LLM integration.}
    \resizebox{1\columnwidth}{!} {
    \begin{tabular}{lc}
        \toprule
        \textbf{Hyperparameter} & \textbf{Value}\\
        \midrule
         Auxiliary Loss Decay & 0.25 \\
         Auxiliary Loss Decay Start & After 32 episodes \\
         Auxiliary Loss Decay Interval & Every 8 episodes \\
         Entropy Increase & \(5e^{-4}\) per auxiliary loss decay \\
         Entropy Decrease & \(2.5e^{-4}\) once teacher's impact reaches 0  \\
         Action Masking Decay & 0.25 \\
         Action Masking Decay Start & After 32 episodes\\
         Action Masking Decay Interval & Every 8 episodes\\
         \bottomrule
    \end{tabular}
    }
    \label{tab:finalHyperparametersLLMIntegration}
\end{table}

\FloatBarrier
\section{Evaluation}
\label{eval}
In this section, we present, analyze, and interpret the results of integrating an LLM to enhance decision-making in ACO. Specifically, we cover:
\begin{enumerate}[label=\textit{\Alph*.}]
    \item Selecting the LLM that produces the most contextually relevant responses for CybORG.
    \item Evaluating combinations of teacher-guided techniques.
    \item Assessing the LLM's impact on the agent's training.
    \item Summarizing the key findings of our study.
\end{enumerate}

\subsection{LLM Evaluation}
As discussed in Section \ref{meth:llmselection}, we evaluated the four LLMs outlined in Table \ref{tab:llmsEvaluatedPhase1} against a dataset of 100 questions using the prompt format shown in Fig. \ref{fig:initialPromptDesign}. To improve the contextual relevance of the LLMs' responses, we formatted CybORG's raw output into a condensed JSON and sentence format. We compare the results using the raw versus transformed inputs in Appendix B of the online supplemental material.

We used BERTScore and manual validation to rank the LLMs' responses. While BERTScore provided useful signals, we observed cases where precision was not proportional to the actual quality of the LLM's response. Table \ref{tab:deficienciesOfBertscore} demonstrates that BERTScore may assign high precision scores to suboptimal recommendations. Row 1 yields a notably higher score than Row 2 despite ignoring a compromised operational server. In contrast, Row 2 recommends a contextually valid alternative to the manually selected answer (patch instead of analyze for User1), but receives a 12.81\% lower precision score. 

Manual scoring using a 0/0.5/1 rubric was ultimately used to select the best model, where the score was:

\begin{itemize}
    \item \textit{0} if the response was irrelevant or failed to produce an extractable answer.
    \item \textit{0.5} if the response contained the correct action or host, or was partially relevant to the scenario.
    \item \textit{1} if the response matched the labeled answer or provided an equivalently valid recommendation. For example, if two equal-priority hosts were exhibiting similar behavior and the LLM recommended the same action on the host not included in the answer.
\end{itemize}

Table \ref{tab:manualLLMScores} summarizes the results and shows that Cyber-Risk-Llama-8B achieved the highest overall score, and was therefore selected as the LLM for our study \cite{huggingface_cyber-risk-llama}.

\subsection{Teacher-Guided Techniques}
We compared the three teacher-guided combinations introduced in Section \ref{meth:teacherguidedtechniques} to identify the most effective technique for downstream LLM integration. 

\subsubsection{Method Summary}
The following combinations were evaluated:
\begin{itemize}
    \item \textit{Feature Space Modification with Reward Shaping.} We appended the teacher's recommendation as a one-hot encoded vector and gradually decayed the magnitude of the reward at each training interval, using the same parameters as in \cite{thollteacherguidedevaluation2025}. Specifically, we added a reward of \(c_{1}\)=2.5 during the first training interval if the agent selected the teacher's recommended action, and a reward of \(c_{1}\)=1.0 if the agent selected an action involving the same host recommended by the teacher, decreasing the reward magnitude by 10\% at every training interval (see \eqref{eq:featureandrewardshaping}).

    \item \textit{Action Masking with Feature Space Modification}. Rather than rewarding the RL agent for following the teacher's guidance, we constrained the agent to select the recommended actions, encouraging it to learn a mapping between the one-hot encoded guidance and an executable action. For the action masking component, we applied a mask that decayed in strength by 25\% every training interval, starting after episode 32 (see \eqref{eq:featureandactionmasking}). 

    \item \textit{Action Masking with Auxiliary Loss.} We combined the direct impact of masking with the indirect effect of an auxiliary loss signal, to have the agent modify its policy to match the sampled actions. We decayed both the auxiliary loss signal and the masking by 25\% every training interval, starting after episode 8 (see \eqref{eq:actionmaskingandauxloss}).
\end{itemize}

\begin{table}
    \centering
    \caption[Deficiencies of Relying Solely on BERTScore.]{Limitations of relying solely on BERTScore for evaluating LLM responses.}
    \resizebox{\columnwidth}{!} {
      \begin{tabular}{llc}
        \toprule
        \textbf{Best Action (Validation Label)} & \textbf{LLM Response} & \textbf{Precision} \\
        \midrule
        Restore Operation1 & Allow Enterprise1 80 & 0.8675 \\
        Analyse User1 & Patch Host: User1 \{'Action': 'Patch', 'Host': 'User1'\} & 0.7564 \\
        \bottomrule
      \end{tabular}
    }
    \label{tab:deficienciesOfBertscore}
\end{table}

\begin{table}
\centering
\caption[LLM Selection - Manual Validation.]{Manual scoring of each LLM across 20 easy, 40 medium, and 40 hard questions in both JSON and sentence formats. The highest scores per row are \textbf{bolded}.}
\resizebox{\columnwidth}{!}{%
\small
  \begin{tabular}{llcccc}
    \toprule
    & & \multicolumn{4}{c}{\textbf{LLMs}} \\
    \cmidrule(lr){3-6}
    & & \textbf{Cyberdost2B} & \textbf{Z7sec} & \textbf{Llama8B} & \textbf{Lily7B} \\
    \midrule
    \multirow{2}{*}{\textbf{Easy (JSON)}} 
      & Total   &  8   & 4.5   & \textbf{12}   & 9.5 \\
      & Average & 0.40 & 0.225 & \textbf{0.60} & 0.475 \\
    \midrule
    \multirow{2}{*}{\textbf{Medium (JSON)}} 
      & Total   & 7 & 10 & \textbf{19.5} & 14 \\
      & Average & 0.175 & 0.250 & \textbf{0.4875} & 0.350 \\
    \midrule
    \multirow{2}{*}{\textbf{Hard (JSON)}} 
      & Total   & 4 & 7.5 & \textbf{14} & 12 \\
      & Average & 0.10 & 0.1875 & \textbf{0.35} & 0.30 \\
    \midrule
    \multirow{2}{*}{\textbf{Easy (Sentence)}} 
      & Total   & 1 & 6 & 6 & \textbf{9} \\
      & Average & 0.05 & 0.30 & 0.30 & \textbf{0.45} \\
    \midrule
    \multirow{2}{*}{\textbf{Medium (Sentence)}} 
      & Total   & 2.5 & \textbf{16} & 8 & 13.5 \\
      & Average & 0.0625 & \textbf{0.40} & 0.20 & 0.3375 \\
    \midrule
    \multirow{2}{*}{\textbf{Hard (Sentence)}} 
      & Total   & 1 & \textbf{13} & 4 & \textbf{13} \\
      & Average & 0.025 & \textbf{0.325} & 0.10 & \textbf{0.325} \\
    \midrule
    \multirow{2}{*}{\textbf{Total (JSON)}} 
      & Total   & 19 & 22 & \textbf{45.5} & 35.5 \\
      & Average & 0.19 & 0.22 & \textbf{0.455} & 0.355 \\
    \midrule
    \multirow{2}{*}{\textbf{Total (Sentence)}} 
      & Total   & 4.5 & 35 & 18 & \textbf{35.5} \\
      & Average & 0.045 & 0.35 & 0.18 & \textbf{0.355} \\
    \bottomrule
  \end{tabular}
}
\label{tab:manualLLMScores}
\end{table}

\subsubsection{Performance Analysis}
In both combinations that use feature space modification, LIME shows the agent exhibits no evidence of being able to map the teacher's recommendation to an executable action. As shown in Tables \ref{tab:rewardShapingAndFeatureSpaceAsOneHotEncoded} and \ref{tab:actionMaskingAndFeatureSpaceAsOneHotEncoded}, whenever the teacher's recommendation appears in the top four actions, its corresponding feature has a low impact on the agent's decision. For clarity, we show only the rank of the one-hot encoded feature (the feature set to 1), where a rank of 132 represents the lowest importance (i.e., least impact on the agent's decision), and 1 the highest.

\begin{table}[!b]
\centering
\caption[LIME Results with Reward Shaping - One-Hot Encoded.]{LIME results for feature space modification with reward shaping. Only the weight of the one-hot encoded feature is shown. 'Reco in Top 4' indicates whether the teacher's recommendation is within the top four actions of the RL agent's policy, with its associated ranking.}
\small
\resizebox{1\columnwidth}{!} {
\begin{tabular}{ccccc}
\toprule
\textbf{Episode} & \textbf{Weight} & \textbf{Ranking} & \textbf{Direction} & \textbf{Reco in Top 4} \\
\midrule
1   & 8.50E-06 & 31 & Towards & No    \\
8   & -7.27E-04 & 22 & Away    & No    \\
16  & 4.14E-04 & 65 & Towards & No    \\
50  & 4.66E-04 & 73 & Towards & No    \\
100 & 1.44E-02 & 30 & Towards & No    \\
200 & 1.14E-01 & 4  & Towards & No    \\
300 & 2.22E-02 & 60 & Towards & Yes/2 \\
500 & 1.19E-01 & 9  & Towards & No    \\
\bottomrule
\end{tabular}
}
\label{tab:rewardShapingAndFeatureSpaceAsOneHotEncoded}
\end{table}

\begin{table}[!b]
\centering
\caption[LIME Results with Action Masking.]{LIME results for feature space modification with action masking. Only the weight of the one-hot encoded feature is shown. 'Reco in Top 4' indicates whether the teacher's recommendation is within the top four actions of the RL agent's policy, with its associated ranking.}
\small
\resizebox{1\columnwidth}{!} {
\begin{tabular}{ccccc}
\toprule
\textbf{Episode} & \textbf{Weight} & \textbf{Ranking} & \textbf{Direction} & \textbf{Reco in Top 4} \\
\midrule
1   & 7.22E-07  & 82 & Towards & Yes/3 \\
8   & 1.62E-03  & 3  & Towards & No    \\
16  & 9.66E-04  & 46 & Towards & No    \\
50  & 2.40E-03  & 41 & Towards & No    \\
100 & -1.30E-02 & 27 & Away    & No    \\
200 & 2.96E-02  & 23 & Towards & No    \\
300 & 2.32E-02 & 52 & Towards & No    \\
500 & -5.98E-03 & 78 & Away    & No    \\
\bottomrule
\end{tabular}
}
\label{tab:actionMaskingAndFeatureSpaceAsOneHotEncoded}
\end{table}

\begin{figure}
    \centering
    \includegraphics[width=1.0\columnwidth]{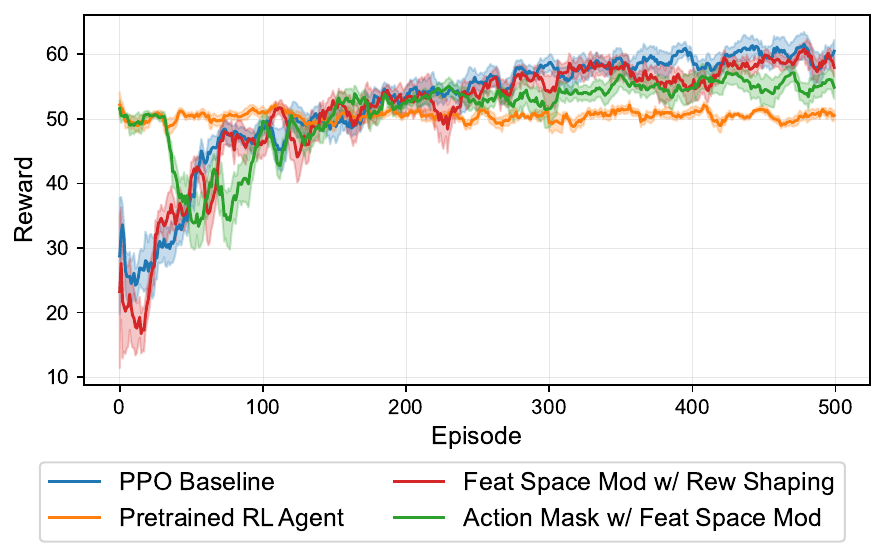}
    \caption[Feature Space Modification Combination]{Comparison of feature space modification combinations against the PPO baseline across 10 independent runs. Shaded regions represent a ±1 SE (using the running average).}
    \label{fig:featureSpaceCombinations}
\end{figure}

\begin{figure}
    \centering
    \includegraphics[width=1.0\columnwidth]{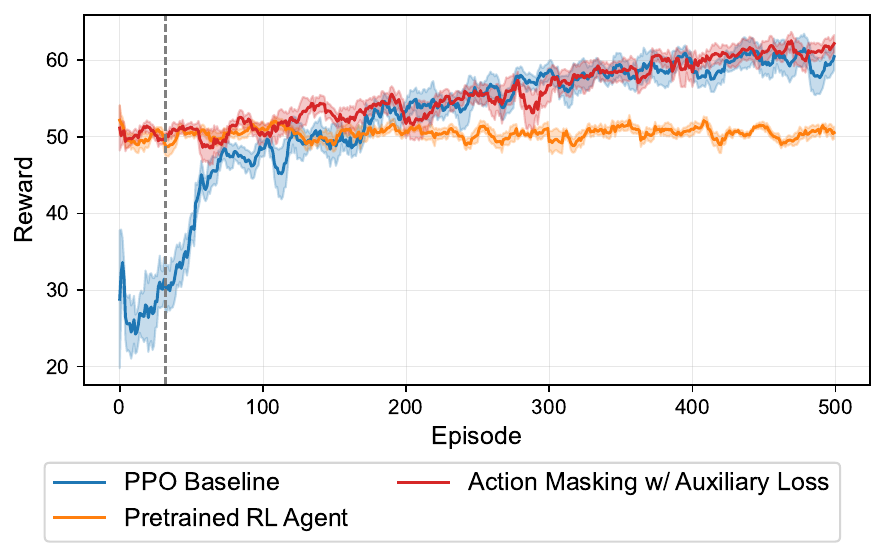}
    \caption[Auxiliary Loss with Action Masking.]{Comparison of action masking with auxiliary loss against the PPO baseline across 10 independent runs. The vertical dashed line indicates the point at which the teacher guidance is completely removed after having been decayed by 25\% per training interval. Shaded regions represent a ±1 SE.}
    \label{fig:ActionMaskingAndAuxiliaryLoss}
\end{figure}

We present the performance of each teacher-guided combination in Figs. \ref{fig:featureSpaceCombinations} and \ref{fig:ActionMaskingAndAuxiliaryLoss}. Only action masking coupled with an auxiliary loss signal exhibited early-stage improvement with a smooth transition to independent RL. After this transition, the guided agent's performance remains superior to the baseline; however, the baseline catches up around episode \(\approx\)170. In contrast, action masking with feature space modification, shows the same initial improvement, but quickly deteriorates once guidance is removed - even falling below the performance of the baseline.

Based on its superior performance and smooth transition to independent RL, we selected action masking with an auxiliary loss signal as our teacher-guided technique to incorporate the LLM into the RL pipeline.

\subsection{LLM Integration}
\label{eval:llintegration}
After selecting the best-performing teacher-guided technique with respect to training efficiency, we replaced the pretrained RL agent with Cyber-Risk-Llama8B, using the methods described in Section \ref{meth:llmintegration} to map its textual output to an executable action \cite{huggingface_cyber-risk-llama}.

\subsubsection{Prompt Design}
Before integrating the LLM into the pipeline, we further refined the prompt developed in Phase 1 during LLM selection.

To begin, we substituted CybORG-specific terms for actions and hosts with generic labels to mitigate biases introuded by Cyber-Risk-Llama8B's training data. Despite this modification, we observed that the LLM consistently prioritized hosts that appeared earlier in the prompt. This behavior likely stems from the model's masked self-attention mechanism, which uses only preceding tokens to compute attention scores - potentially assigning disproportionate weight to earlier tokens \cite{vaswani_attention_2017}. To address this, we ordered the hosts in the prompt by their respective priorities.

To further enhance performance, we added the minimum number of hops each host had from the operational server to the prompt - but restricted this to hosts on critical paths (i.e., the shortest attack paths an adversary could follow to propagate its presence to the operational server).

\begin{figure}
    \centering
    \includegraphics[width=1.0\columnwidth]{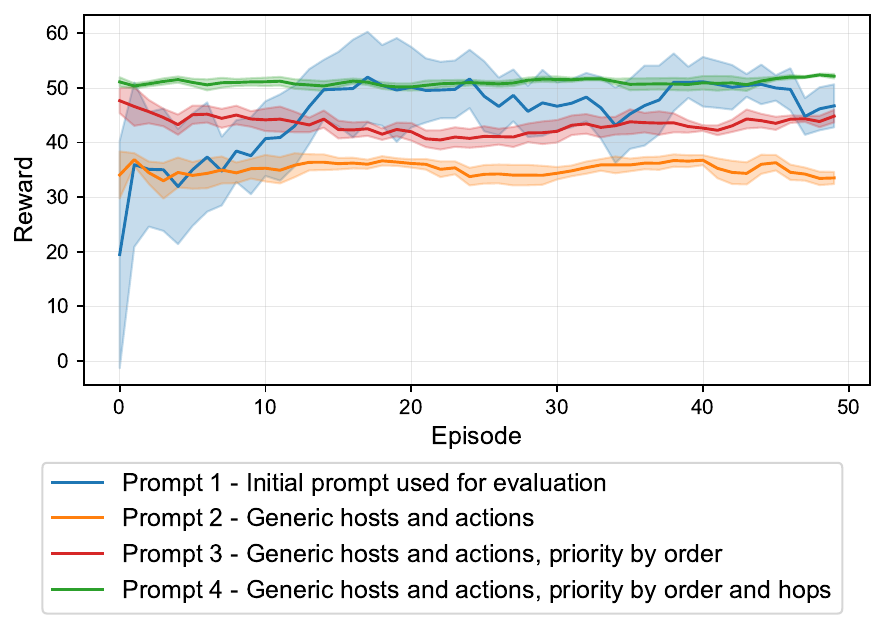}
    \caption[LLM Performance with Different Prompts.]{Evaluation of Cyber-Risk-Llama8B using various prompts across 10 independent runs over 50 episodes. Each curve shows the per-episode mean reward with shaded regions representing a ±1 standard error.}
    \label{fig:PromptEvaluation}
\end{figure}

Fig. \ref{fig:PromptEvaluation} illustrates the performance of the four prompt formats we tested for the LLM integration. Prompt 4, denoted by the green curve, was ultimately used for our study.

\subsubsection{LLM-Guided Training vs PPO Baseline}
\begin{figure}
    \centering
    \includegraphics[width=1.0\columnwidth]{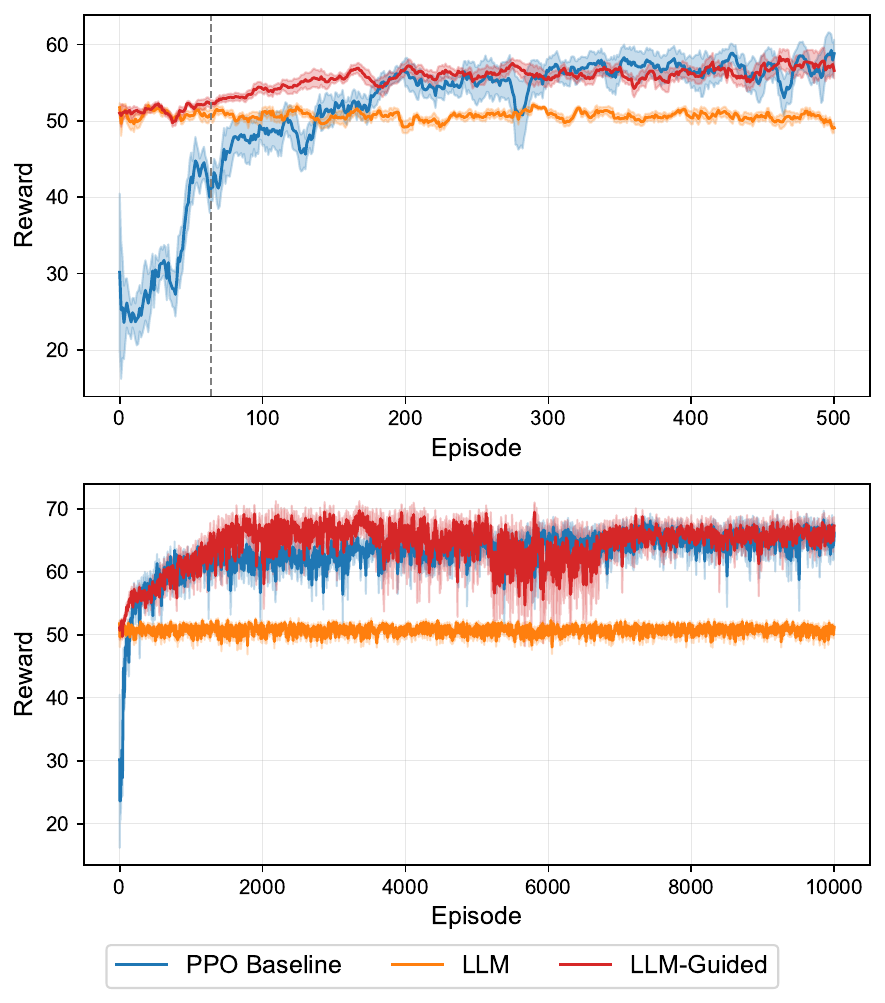}
    \caption[LLM-Guided Performance.]{Comparison of LLM-guided training using action masking with auxiliary loss against the PPO baseline across 10 independent runs. The vertical dashed line marks the point at which the agent has transitioned to fully independent RL (i.e., learning solely from the environment's signals). For clarity, the dashed line is only shown on the top plot.
    Top: mean reward after applying a 10-episode running average with a ±1 standard error for 500 episodes.
    Bottom: mean reward after applying a 10-episode running average with a ±1 standard error for 10,000 episodes.}
    \label{fig:LLMAuxLossAndActionMask}
\end{figure}

We present the comparison of our LLM integration using the gradually decaying, inference-only action masking and auxiliary loss signal against the baseline in Fig. \ref{fig:LLMAuxLossAndActionMask}. 

Integrating the LLM as a teacher accelerates convergence to a favorable policy and enables continued learning from a higher initial position. However, the guided agent exhibits notably higher variance, particularly between episodes \(\approx\)5,200 and \(\approx\)6,600. 

%Putting a bit farther down so comes after referenced

A likely source of this instability is how we incorporated the teacher's guidance, which may lead the RL agent to converge on an unreliable policy. Table \ref{tab:PretrainedVsDistilledProbs} compares the action distributions of an agent trained solely on the teacher's feedback for 80 episodes to agents trained independently for 100 and 10,000 episodes, respectively. The LLM-guided agent exhibits extreme confidence in a single action, with its highest probability exceeding the second-highest by a factor of approximately 3,310. In contrast, independently trained agents show more balanced policies, with the top two probabilities differing by only a factor of approximately 2.5. 

The highly peaked distribution that the teacher-guided agent converges to disrupts learning beyond the LLM's baseline. Small perturbations to the logits of the most probable action can cause disproportionate shifts in the resulting softmax, greatly affecting the probabilities of alternate actions. While the LLM expedited early training, it introduced instability by reinforcing a narrow, overconfident policy during the guided phase.

\subsubsection{LLM Guidance as a Distribution}
A potential remedy to the discussed limitation was to have the LLM output a distribution over actions instead of a single recommendation. This would have ensured the policy remains appropriately distributed during the teacher-guided phase, facilitating a smoother transition to independent PPO. Such a distribution could be obtained by querying the LLM individually for each action, as demonstrated by Z. Zhou et al. \cite{zhou_llm4rl_2024}. However, due to the LLM's computational overhead and CybORG's large action space, we deemed this approach infeasible.

\begin{figure}
    \centering
    \includegraphics[width=1.0\columnwidth]{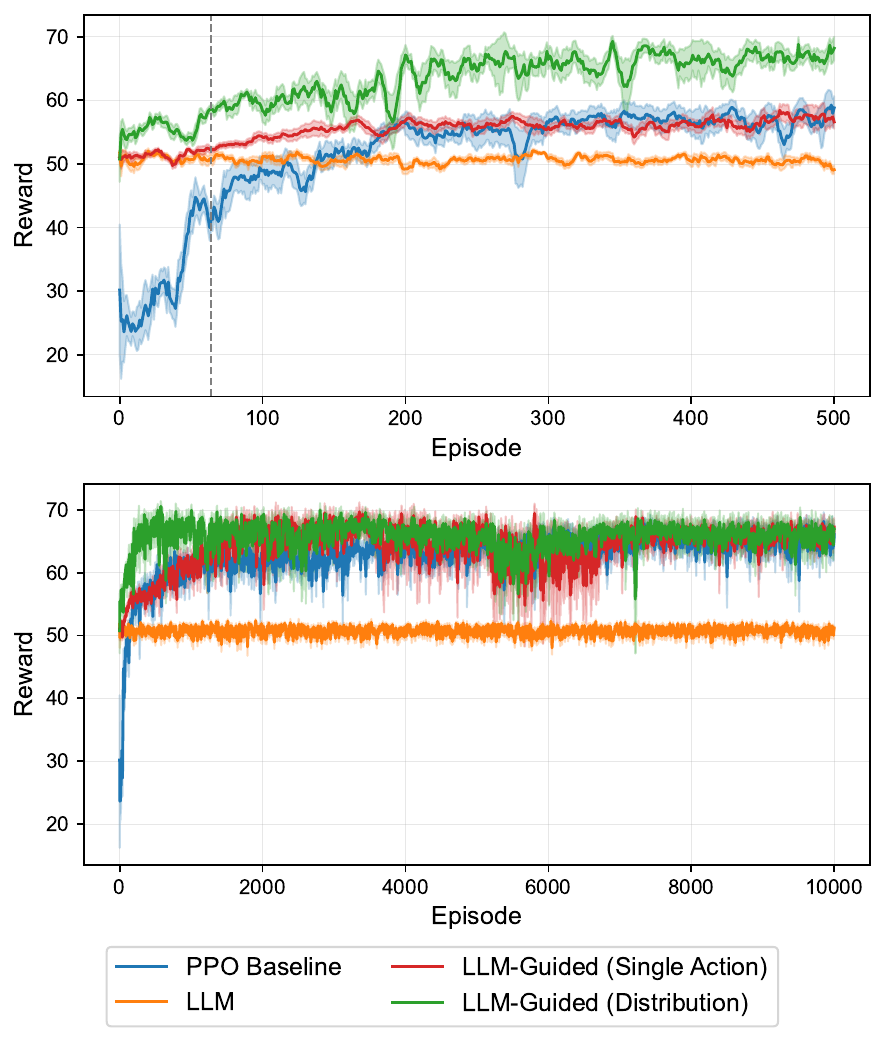}
    \caption[LLM-Guided Performance Using a Distribution.]{Demonstrating the performance of LLM integration using a distribution-based approach. The vertical dashed line indicates the point at which the teacher-guided agents have transitioned to fully independent RL (i.e., learning solely from the environment's signals). For clarity, the dashed line is only shown on the top plot.    
    Top: mean reward after applying a 10-episode running average with a ±1 standard error for 500 episodes. 
    Bottom: mean reward after applying a 10-episode running average with a ±1 standard error for 10,000 episodes.}
    \label{fig:distvssingleaction}
\end{figure}

Instead, we explored mapping the LLM's output into a probability distribution aligned with CybORG's action space. Fig. \ref{fig:distvssingleaction} compares incorporating the LLM's guidance as a single action with using a distribution; both implementations share identical hyperparameters.

%Putting Figure here so doesn't break the conclusion

\begin{table}
    \centering
    \caption[Comparing Probability Distributions with LLM-Guided Agent.]{Comparison of the top three action probabilities for the LLM-guided agent, and agents trained independently for 100 and 10,000 episodes. Column titles are abbreviated for readability.}
    \resizebox{1\columnwidth}{!}{
    \begin{tabular}{lccc}
        \toprule
        & \textbf{LLM-Guided} & \textbf{100 Eps RL} & \textbf{10,000 Eps RL}\\
        \midrule
         Largest Prob & 99.96\% & 21.64\% & 54.86\% \\
         Second Largest Prob &  0.03\% & 8.53\% & 21.91\% \\
         Third Largest Prob & 0.01\% & 5.52\% & 8.86\% \\
         \bottomrule
    \end{tabular}
    }
    \label{tab:PretrainedVsDistilledProbs}
\end{table}

Incorporating the LLM's guidance as a distribution yields greater training efficiency than the single-action recommendation after the transition to independent RL at episode 64, achieving an average reward of \(\approx\)68 instead of \(\approx\)59. The distribution-guided agent peaks quickly and eventually stabilizes with the baseline at approximately episode 4,800, demonstrating the ability to converge on a favorable policy \(\approx\)4,500 episodes faster. 

Notably, the distribution-guided agent also outperforms the single-action approach during the masking period. This improvement results from sampling from the LLM's policy rather than greedily selecting the highest-probability action, highlighting the potential benefits of stochastic sampling over deterministic selection. Our guided agent yields over 2x higher rewards starting and continuing to learn from an average reward of \(\approx\)55 instead of \(\approx\)25.

We describe the procedure for mapping the LLM's output into a probability distribution and using this to guide training in Appendix C of the online supplemental material. 

\subsection{Summary of Key Findings}
In summary, we found that combining an auxiliary loss signal with action masking was the most effective teacher-guided technique. Prompt engineering played a critical role in improving the LLM's output quality. Finally, incorporating the LLM's guidance as an entire distribution rather than a single-action improved the effectiveness of our solution, most notably during the transition to independent RL.

Overall, our LLM-guided pipeline enables RL agents to:
\begin{itemize}
    \item Begin learning from a higher baseline, allowing them to progress without executing obviously undesirable actions.
    \item Converge to a favorable policy in fewer timesteps than the baseline agent. The final policy performance is comparable to agents trained solely through independent RL.
\end{itemize}

%Make it so if there is only a few lines, then start at next page
\needspace{7\baselineskip}
\section{Conclusion}
\label{conc}
ACO aims to train agents capable of executing effective actions with minimal human oversight. However, current ACO applications require these agents to learn from scratch, often resulting in clearly suboptimal behavior in training as they explore unfavorable actions to learn their consequences. In this study, we address this limitation by incorporating external knowledge in the form of an LLM, enabling the agent to bypass such undesirable behavior and accelerate learning.

\subsection{Contributions}
Our primary contribution is demonstrating that LLMs, when properly prompted, can significantly improve training for RL agents in ACO. Using CybORG's Cage Challenge 2 as a proof of concept, we show that LLMs can accelerate learning and improve the initial policy performance of agents.

In addition to illustrating the positive impact of LLMs in ACO, we provide:
\begin{itemize}
    \item A framework for efficiently evaluating and selecting the most effective LLM for a target environment.
    \item A teacher-guided technique that combines action masking with an auxiliary loss signal to support both early-stage guidance and a smooth transition to independent RL.
    \item A pipeline for incorporating an LLM into the decision-making process of RL agents without modifying the LLM's parameters or the environment.
\end{itemize}

\subsection{Limitations}
While our study presents meaningful contributions to ACO, several limitations exist across the different phases.

\textit{Selecting an LLM.}
While we attempted an autonomous, objective selection process for Phase 1, our final scoring of the LLMs relied on manual validation of response quality. This inherent subjectivity and the use of discrete scores (instead of scoring across a spectrum) limited the quality of this evaluation.

\textit{Environment Limitations.}
CybORG, while one of the most impressive ACO environments, simplifies the complex state space of real networks, under-represents user activity, and forces adversaries and defenders to execute actions in an unrealistic, sequential fashion. Although our study provides a strong proof of concept for LLM integration in ACO, these inherent limitations of using a simulated environment prevent us from guaranteeing that comparable results would hold in a real-world operational setting.

\textit{Parsing CybORG's Output.} Although our LLM integration aimed to reduce feature engineering, we still formatted CybORG's raw state space into a condensed form to improve the LLM's response quality. In doing so, we omitted potentially relevant features - such as a host's operating system type and architecture - that may have affected performance.

\textit{Final Prompt Design.} To ensure clarity and LLM alignment, we included CybORG-specific action definitions in the prompt. While this improved response quality, it may have undermined Phase 1's assessment of each LLM's baseline cybersecurity knowledge by injecting domain-specific context.

\textit{LLM Resource Requirements.} Cyber-Risk-Llama8B requires 16 GiB of memory to load and approximately 3.69 seconds to generate a non-cached response on a DGX-H100 GPU, compared to the millisecond-scale inference time of the PPO agent. This substantial computational cost raises potential scalability concerns with our solution in resource-constrained environments.

\subsection{Future Work}
Although our study has limitations, its novel contributions lay the groundwork for significant advancements in ACO. These opportunities include:
\begin{itemize}
    \item Improving the integration technique to enable a smoother transition from teacher-guided to independent RL. 
    \item Integrating the LLM into the emerging multi-agent RL field.
    \item Leveraging the LLM to augment decision-making for the red agent.
    \item Fine-tuning the LLM for the CybORG environment to improve alignment.
    \item Incorporating the LLM into the feature engineering process to enhance state representation.
    \item Applying the LLM in an emulated (rather than simulated) environment to test in a more realistic environment.
    \item Exploring encoder-only models as an alternative to decoder-only architectures for bidirectional pattern recognition and to reduce computational overhead.
    \item Defending the LLM in an adversarial setting to ensure robustness against targeted attacks.
\end{itemize}

Integrating an LLM into the RL pipeline to augment decision-making represents a key milestone in ACO applications, creating a strong foundation for future advancements in autonomous cybersecurity.

\bibliographystyle{unsrt}
\bibliography{skeleton}

\appendix 

\section{LLM Prompt}
\label{app:llmPrompt}
We provide the prompt used to evaluate the LLM integration in Listing \ref{lst:LLMPrompt}.
\begin{lstlisting}[basicstyle=\ttfamily\footnotesize,caption={Final prompt provided to the LLM for our study. Extra newlines have been added to the original to be displayed in the width of a column.},label={lst:LLMPrompt}, breaklines=true, breakatwhitespace=true]

**Role**:
You are an expert at cybersecurity and network defense.

**Objective**:
You must choose **ONE ACTION** to execute on the most critical host to maintain network functionality.

**Prioritize Hosts by Distance to the Target**
- Prioritize hosts that are closest to the operational server:
- host8: 0 hops.
- host4: 1 hops.
- host2: 2 hops.
- host3: 2 hops.
- host10: 3 hops.
- host11: 3 hops.
- host12: 3 hops.
- host13: 3 hops.

**Determine the Best Action**
For the selected host, choose the best action based on the following definitions:
- **action1**: Use to gather additional information on a host, such as active processes, connections, or files.
- **action2**: Use to revert a fully compromised host where the attacker has gained persistence. 
- **action3**: Use to terminate non-persistent malicious processes before they escalate privileges or gain persistence.
- **action4**:Use to apply security updates and reduce the likelihood of future exploitation.
- **action5**: Use to block network access to and from a compromised host to prevent lateral movement.
- **action6**: Use to restore network access to a previously isolated host once it is deemed safe.

### **Current Network State**:
host8| IP: 10.0.179.17, NOT ISOLATED, Files: [], Processes: [11 processes with: (Remote IP: 10.0.243.220 and Port: 22)], Scans: []
host4| IP: 10.0.232.83, NOT ISOLATED, Files: [cmd.exe at C:\temp\ (Density: 0.9, Signed: No), escalate.exe at C:\temp\ (Density: 0.9, Signed: No)], Processes: [1 process with: (Remote IP: 10.0.243.220 and Port: 57984)], Scans: []
host2| IP: 10.0.232.84, NOT ISOLATED, Files: [escalate.sh at /tmp/ (Density: 0.9, Signed: No)], Processes: [11 processes with: (Remote IP: 10.0.243.220 and Port: 22)], Scans: []
host3| IP: 10.0.232.93, NOT ISOLATED, Files: [], Processes: [], Scans: []
host10| IP: 10.0.243.219, NOT ISOLATED, Files: [], Processes: [], Scans: []
host11| IP: 10.0.243.211, NOT ISOLATED, Files: [], Processes: [], Scans: [1 scan with: (Remote IP: 10.0.243.220 and Port: 445), 1 scan with: (Remote IP: 10.0.243.220 and Port: 139), 1 scan with: (Remote IP: 10.0.243.220 and Port: 135), 1 scan with: (Remote IP: 10.0.243.220 and Port: 3389)]
host12| IP: 10.0.243.213, NOT ISOLATED, Files: [], Processes: [], Scans: []
host13| IP: 10.0.243.215, NOT ISOLATED, Files: [cmd.sh at /tmp/ (Density: 0.9, Signed: No)], Processes: [1 process with: (Remote IP: 10.0.243.220 and Port: 54437)], Scans: []
host1| IP: 10.0.232.87, NOT ISOLATED, Files: [], Processes: [], Scans: []
host5| IP: 10.0.179.27, NOT ISOLATED, Files: [], Processes: [], Scans: []
host6| IP: 10.0.179.20, NOT ISOLATED, Files: [], Processes: [], Scans: []
host7| IP: 10.0.179.18, NOT ISOLATED, Files: [], Processes: [], Scans: []
host9| IP: 10.0.243.220, NOT ISOLATED, Files: [], Processes: [], Scans: []

Your response must follow this format exactly: <action> <hostname> (e.g., actionx hostx).

Best Action:
\end{lstlisting}
%Make it so if there is only a few lines, then start at next page
\needspace{7\baselineskip}
\section{Parsing CybORG's Output}
\label{app:parsingCybORGOutput}
\begin{figure*}
    \centering
    \includegraphics[width=\columnwidth]{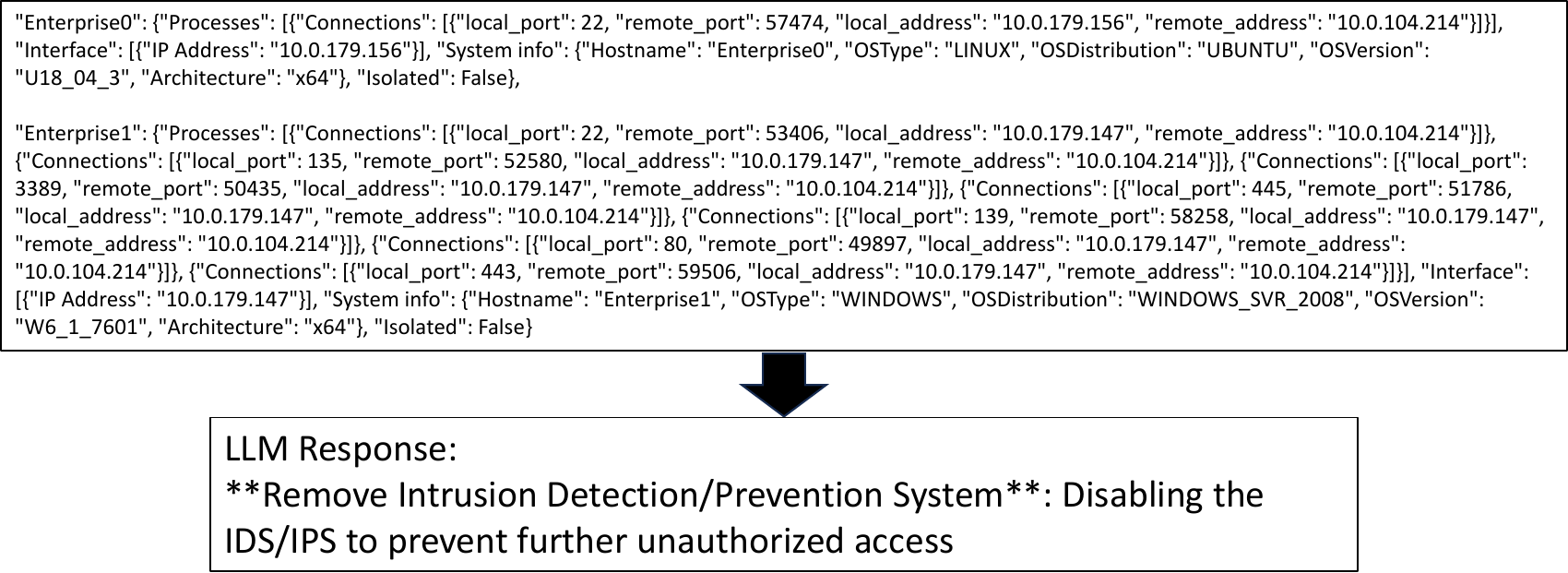}
    \vspace{1em} % space between subfigures

    {\scriptsize\textbf{(b) LLM output using parsed CybORG state}\par}
    \includegraphics[width=\columnwidth]{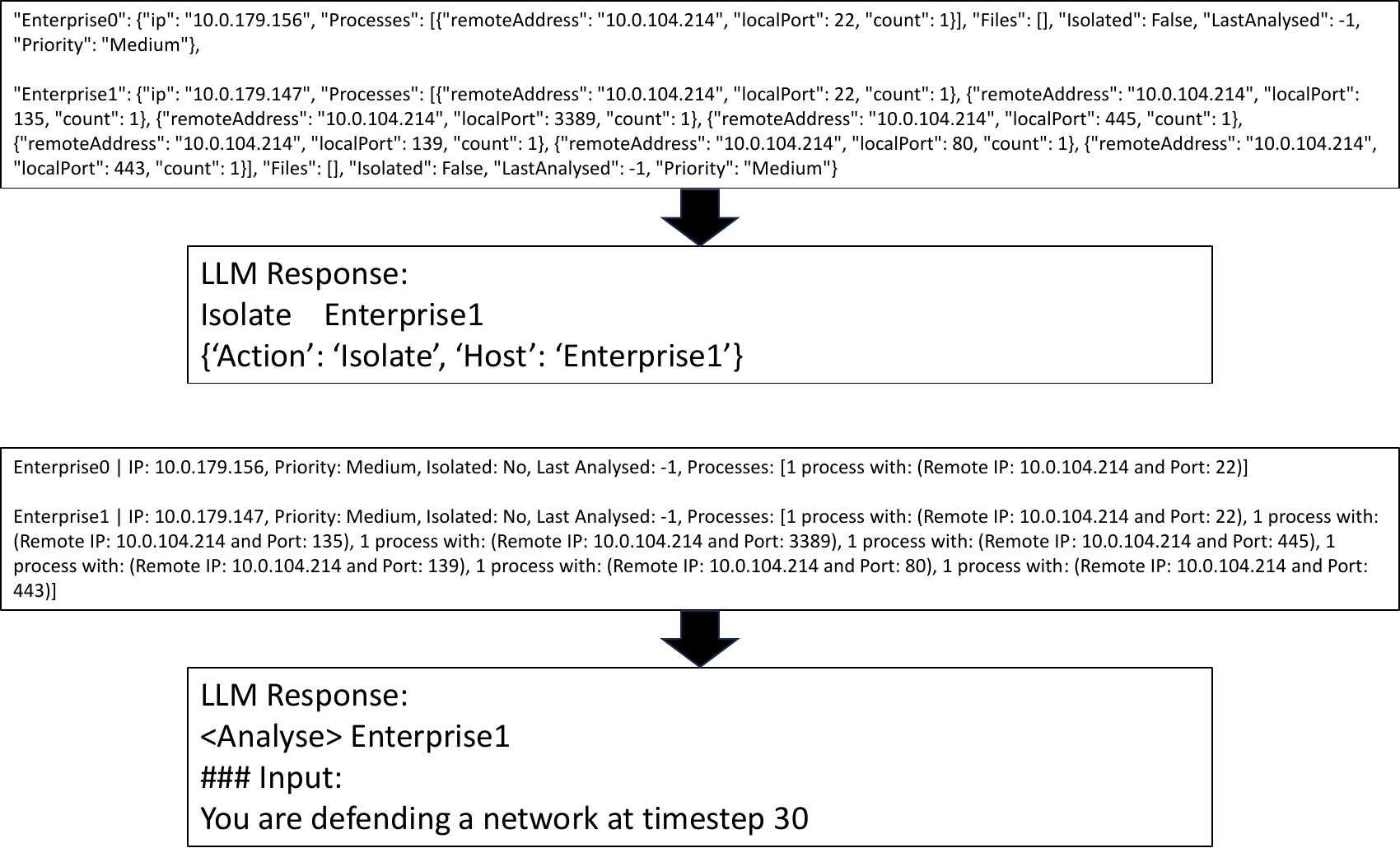}

    \caption[LLM response to raw and condensed CybORG state]{
        LLM response to the raw and condensed CybORG state representations. For clarity, only 2 of the 13 hosts' states are shown.
        Top: response generated from the raw (unfiltered) CybORG output.
        Bottom: response generated from the condensed CybORG output.
    }
    \label{fig:LLMparsedvsraw}
\end{figure*}

We observed that directly providing CybORG's raw output to the LLM produced unsatisfactory responses. To mitigate this, we converted the raw state into more compact JSON and sentence-based formats. We compare the LLM's output across the raw and condensed inputs in Fig. \ref{fig:LLMparsedvsraw}. The raw state produces a response from which no valid action can be reliably extracted, whereas the condensed formats yield output that can be consistently mapped to valid actions.

\section{Mapping the LLM to a Distribution}
\label{app:extractingProbsFromLLM}
Our original strategy of incorporating the teacher's guidance through an auxiliary loss signal enhanced training performance, but had a critical limitation. Recommending a single action caused the agent to converge to a peaked policy, suppressing the likelihood of all other actions. Consequently, further improvements became impossible until the distribution flattened, allowing the baseline (independently trained agent) to match, or even surpass the teacher-guided agent's performance. 

To address this limitation, we mapped the LLM's output to a full probability distribution and used this entire distribution - instead of a single action - to compute the auxiliary loss signal to guide the agent's learning. This allowed the agent to learn a policy over the entire action space instead of greatly inflating a single action probability.

We first identified the tokens corresponding to the LLM's recommended action and host, then extracted the softmax distributions from which they were sampled. Once extracted, we parsed the probabilities for tokens corresponding to all valid actions and hosts, discarding tokens unrelated to CybORG. 

We then applied Cartesian multiplication between the action and host distributions to compute the probability of executing every action on every host, aligning with CybORG's action space. Finally, we normalized this combined distribution and used it as the auxiliary loss signal to guide training.

\begin{figure*}
    \centering
    \includegraphics[scale=0.75]{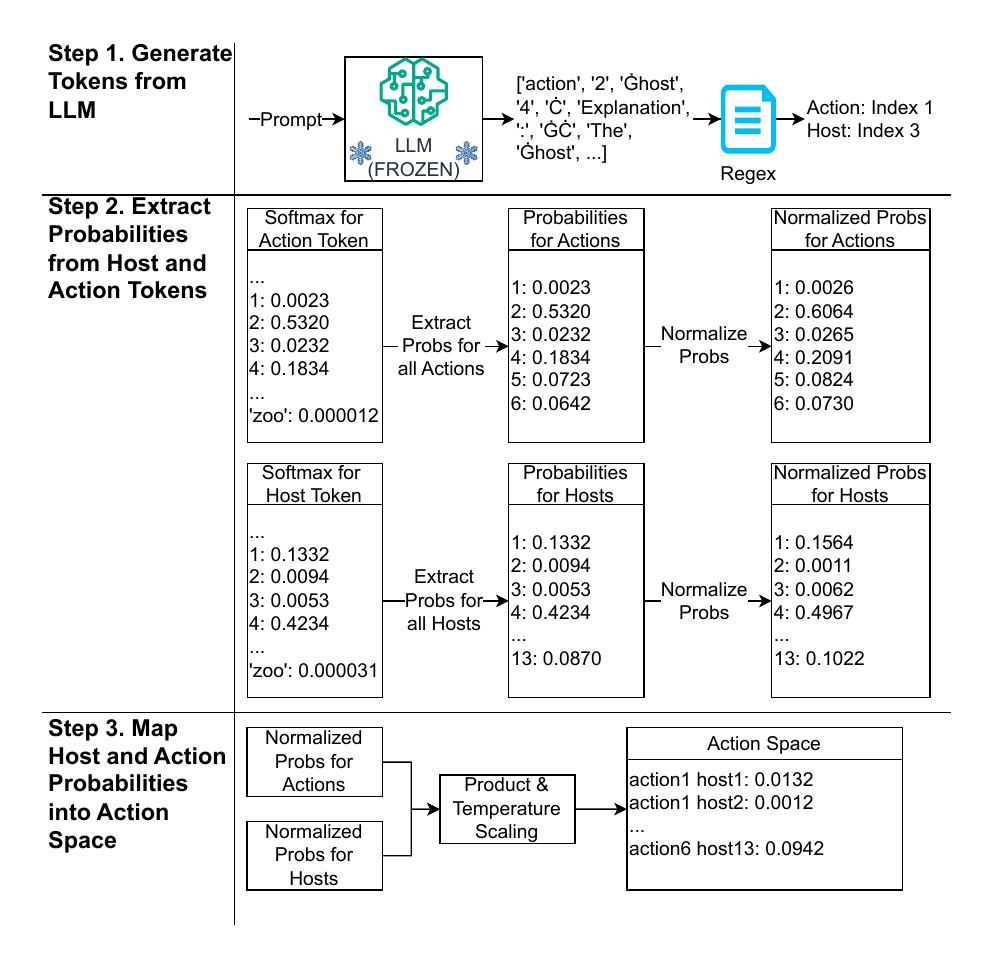}
    \caption[Extracting Probability Distribution from LLM.]{Extracting a probability distribution from the LLM. 
    In Step 1, we parse the tokens generated by the LLM and extract those representing the recommended action and host.
    In Step 2, the softmax distributions for the LLM's predicted action and host are used to derive probabilities for every valid action and host in the environment. These are then normalized to form probability distributions scoped to CybORG. The `zoo' token is meant to illustrate that the LLM outputs a distribution across every possible token.
    In Step 3, the individual host and action probabilities are combined via Cartesian multiplication to produce a distribution across all action-host pairs (matching CybORG's action space). Although we experimented with temperature scaling to sharpen the distribution (i.e., shift probabilities towards 0 or 1), this was excluded from the final implementation.}
    \label{fig:extractingLLMDistProcess}
\end{figure*}

We illustrate the full procedure for mapping the LLM's output into a probability distribution over CybORG's action space in Fig. \ref{fig:extractingLLMDistProcess}.

The auxiliary loss process we originally used computes the probability of selecting the LLM's single-action recommendation in the agent's current policy; therefore, it cannot be used when the LLM outputs a distribution. As such, we modified the loss term to use Kullback-Leibler (KL) divergence, which computes a scalar that quantifies the difference between two probability distributions \cite{cui_kl_2025}. Because KL divergence captures the entire distribution, it enables the RL agent to align its policy with the LLM's full set of recommendations, rather than optimizing for a single action.

We present the results of incorporating the LLM's guidance as a distribution in Fig. \ref{fig:distvssingleaction} under Section \ref{eval:llintegration}.

\makeatletter
\setlength{\@dblfptop}{0pt}
\makeatother

\end{document}